\documentstyle[15lomcon,cite,epsfig]{article}

\def\roughly#1{\mathrel{\raise.3ex\hbox
{$#1$\kern-.75em\lower1ex\hbox{$\sim$}}}}

\def\gs{\roughly>}

\begin{document}

\title{NEUTRINO EMISSION FROM A STRONGLY MAGNETIZED DEGENERATE ELECTRON GAS: THE COMPTON MECHANISM VIA A NEUTRINO MAGNETIC MOMENT}

\author{A. V. Borisov \footnote{e-mail: borisov@phys.msu.ru},
 B. K. Kerimov \footnote{late}}

\address{Faculty of Physics, Moscow State University, 119991 Moscow, Russia}

\author{P. E. Sizin \footnote{e-mail: mstranger@list.ru}}

\address{Department of Higher Mathematics, Moscow State Mining University,\\119991 Moscow, Russia}


\maketitle\abstracts{~We derive relative upper bounds on the effective magnetic moment of Dirac neutrinos from comparison of the standard weak and electromagnetic mechanisms of the neutrino luminosity due to the Compton-like photoproduction of neutrino pairs  in a degenerate gas of electrons on the lowest Landau level in a strong magnetic field. These bounds are close to the known astrophysical and laboratory ones.}

~~~{\bf 1.} Neutrino emission is the main source of energy losses of stars in the late stages of their evolution \cite{book}. As is well known, neutron stars (NSs) can have strong magnetic fields $H\gs 10^{12}~{\rm G}$, the NSs with $H\sim 10^{14}-10^{16}~{\rm G}$ are called magnetars \cite{DT}.

In this report, we consider one of the main processes of neutrino emission in the outer regions of NSs (for a review of various neutrino processes, see \cite{YKGH}) that is photoproduction of neutrino pairs ($\gamma e \to e\nu\bar{\nu}$) in a degenerate gas of electrons through two mechanisms: the weak one via standard charged and neutral weak currents and the electromagnetic one via neutrino electromagnetic dipole moments arising in extended versions of the Standard Model \cite{book,K} (for a recent review, see \cite{GS}). By comparison of the neutrino luminosities due to these two mechanism, $L_{w}$ and $L_{em}$, we derive relative upper bounds on the neutrino effective magnetic moment (NEMM)
\vspace{-0.3cm}
\begin{equation}
\label{mub}
\bar{\mu}_{\nu} = (\mu_\nu^2 + d_\nu^2)^{1/2},
\vspace{-0.3cm}
\end{equation}
restricting ourselves to the case of Dirac neutrinos. Here $\mu_\nu$ and $d_\nu$ are the neutrino magnetic and electric dipole moments, respectively.

~~~{\bf 2.} We assume that the electron gas is degenerate and strongly magnetized:
\begin{equation}
\label{cond}
T\ll \mu -m,\, H>((\mu/m)^2 - 1)H_0/2,
\end{equation}
where $T$ and $\mu\simeq \mu(T=0)\equiv \varepsilon_{\rm F} = (m^2 + p_{\rm F}^2)^{1/2}$ are the temperature and chemical potential of the gas, $\varepsilon_{\rm F}$ and $p_{\rm F}$ are the Fermi energy and momentum, $H_0 = m^2/e \simeq 4.41\times 10^{13}~{\rm G}$, $m$ and $-e$ are the electron mass and charge (we use the units with $\hbar=c=k_{\rm B}=1$). Under the conditions (\ref{cond}), electrons occupy only the lowest Landau level in the magnetic field with $p_{\rm F} = 2\pi^2 n_e/(eH)$, where $n_e$ is the electron concentration, and the effective photon mass is generated which is equal to the plasmon frequency $\omega_p = ((2\alpha/\pi)(p_{\rm F}/\varepsilon_{\rm F})H/H_0)^{1/2}m$, $\alpha$ is the fine-structure constant.

For the nonrelativistic case, $p_{\rm F}\ll m$ and $\omega_{p}\ll T$, the neutrino luminosities are expressed as follows:
\begin{equation}
\label{LwNR}
L_w = 3.49\times 10^2 H_{13}^{2}\rho_{6}^{-1}T_{8}^{9}~\mbox{erg}\,\mbox{cm}^{-3}\,\mbox{s}^{-1},
\end{equation}
\begin{equation}
L_{em} = 4.06\times 10^{30}(\bar{\mu}_{\nu}/\mu_{\rm B})^{2} \rho_6^{2}T_{8}^{3}~\mbox{erg}\,\mbox{cm}^{-3}\,\mbox{s}^{-1},
\end{equation}
where $H_{13}=H/(10^{13}~{\mbox{G}})$,
$\rho_6=\rho/\left(10^6 ~{\mbox{g}}/{\mbox{cm}^3}\right)$, $T_8=T/(10^8~\mbox{K})$, and $\mu_{\rm B}$ is the Bohr magneton. Note that Eq. (\ref{LwNR}) is in agreement with the result of Ref. \cite{S}. Assuming $L_{em} < L_{w}$, we obtain the upper limit on the NEMM (\ref{mub}): ${\bar \mu }_{\nu}/\mu _{\rm B} < 9.3 \times {10^{ - 15}}{H_{13}}\rho _6^{ - 3/2}T_8^3$, and, for $T= 1.8\times 10^8~{\mbox K},\, H= 2.5\times 10^{12}~{\mbox {G}},\, \rho=5.4\times 10^4~{\mbox{g}}/{\mbox{cm}^3}$, it gives ${\bar \mu }_\nu/\mu_{\rm B} < 1.1 \times 10^{-12}$, which is close to the known astrophysical bounds \cite{PDG}.

For the relativistic case, $p_{\rm F}\gg m$ and $\omega_{p}\gg T$, we obtain
\begin{equation}
\label{LwUR}
L_w = 2.63\times 10^{-2}H_{13}^{43/4}\rho_{6}^{-6}T_{8}^{3/2}\exp(-1.92 H_{13}^{1/2} T_{8}^{-1})~\mbox{erg}\,\mbox{cm}^{-3}\,\mbox{s}^{-1},
\end{equation}
\begin{equation}
\label{LemUR}
L_{em} = 3.02\times 10^{30}(\bar{\mu}_{\nu}/\mu_{\rm B})^{2} H_{13}^{11/4}T_{8}^{3/2}\exp(-1.92 H_{13}^{1/2} T_{8}^{-1})~\mbox{erg}\,\mbox{cm}^{-3}\,\mbox{s}^{-1},
\end{equation}
and a strong relative bound ${\bar \mu }_\nu /\mu _{\rm B} < 9.3\times 10^{-17}H_{13}^{4}\rho_6^{-3} = 7.6\times 10^{-16}$ for $H_{13} = 300, \rho_6= 10^3$.
However, under these conditions, the plasmon decay ($\gamma \to \nu\bar{\nu}$) is a much more effective mechanism of neutrino emission\cite{RCh}. Comparing the corresponding luminosity with that of Eq. (\ref{LemUR}), we derive
a considerably less stringent bound ${\bar \mu }_\nu /\mu _{\rm B} < 1.7\times 10^{-12}H_{13}^{1/2} = 2.9\times 10^{-11}$ (for $H_{13} = 300$), which is close to the conservative bound $\mu_\nu < 0.54\times 10^{-10}\mu _{\rm B}$ \cite{PDG} and the most stringent laboratory limit $\mu_\nu < 3.2\times 10^{-11}\mu _{\rm B}$ \cite{GEMMA}.

We thank P.~A.~Eminov for useful discussions, K.~V.~Stepanyantz and \linebreak O.~\,G.~\,Kharlanov for help in numerical calculations.

\end{document}